\documentclass[submission,copyright,creativecommons]{eptcs}
\providecommand{\event}{MeTRiD 2018} 
\usepackage{underscore}           
\usepackage{graphicx} 
\usepackage{color,soul} 
\usepackage{amsmath}
\usepackage{paralist}
\usepackage{amsthm}
\newtheorem{definition}{Definition}
\newtheorem{examplex}{Example}
\newenvironment{example}
  {\pushQED{\qed}\examplex}
  {\popQED\endexamplex}

\title{Process Network Models for Embedded System Design Based on the Real-Time BIP Execution Engine%
\thanks{The research leading to these results has
received funding from the European Space Agency
project \href{http://www-verimag.imag.fr/MoSATT-CMP.html?lang=en}{MoSaTT-CMP}, Contract No. 4000111814/14/NL/MH}
}
\author{Fotios Gioulekas
\institute{Department of Informatics\\
Aristotle University of Thessaloniki\\
54124 Thessaloniki, Greece}
\email{gioulekas@csd.auth.gr}
\and
Peter Poplavko
\institute{Mentor\textregistered. A Siemens Business\\
110, rue Blaise Pascal\\
Inovallée Montbonnot\\
38334 ST ISMIER CEDEX\\
France}
\email{\quad ppoplavko@gmail.com}
\and
Panagiotis Katsaros
\institute{Department of Informatics\\
Aristotle University of Thessaloniki\\
54124 Thessaloniki, Greece}
\institute{
\and
Information Technologies Institute\\
CERTH, Thessaloniki, Greece
}
\email{katsaros@csd.auth.gr}
\and
Pedro Palomo
\institute{Deimos-Space S.L.U\\
Madrid\\
Spain}
\email{\quad pedro.palomo@deimos-space.com}
}

\def\titlerunning{Process Network Models for Embedded System Design Based on Real-Time BIP}
\def\authorrunning{F. Gioulekas, P. Poplavko, P. Katsaros \& P. Palomo}
\begin{document}
\maketitle

\begin{abstract}
Existing model-based processes for embedded real-time systems support the analysis of various non-functional properties, most notably schedulability, through model checking, simulation or other means. The analysis results are then used for modifying the system's design, so that the expected properties are satisfied. A rigorous model-based design flow differs in that it aims at a system implementation derived from high-level models by applying a sequence of semantics-preserving transformations. Properties established at any design step are preserved throughout the subsequent steps including the executable implementation. We introduce such a design flow using a process network model of computation for application design at a high level, which combines streaming and reactive control processing with task parallelism. The schedulability of the so-called FPPNs (Fixed Priority Process Networks) is well-studied and various solutions have been presented. This article focuses on the design flow's steps for deriving executable implementations on the BIP (Behavior - Interaction - Priority) runtime environment. FPPNs are designed using the TASTE toolset, a convenient architecture description interface. In this way, the developers do not program explicitly low-level real-time OS services and the schedulability properties are guaranteed throughout the design steps by construction. The approach has been validated on the design of a real spacecraft on-board application that has been scheduled for execution on an industrial multicore platform.    
\end{abstract}

\section{Introduction}

The model-based design of embedded real-time systems takes place by modeling the system components and their interactions from the early design stages. An adequate design model includes the application behavior with its partitioning into hardware-software components, the overall system architecture, and the mapping of the application to the system's architecture. Such a high-level model enables the analysis, the early verification of design, and estimations on the system's performance~\cite{BRAU20181}.

The related works are roughly classified into two main classes of approaches. Architecture-centric model-based designs are based on an architecture description, whereas via model transformations the system's non-functional properties are analyzed with appropriate tools~\cite{Hugues:2008}. Schedulability depends on certain assumptions for the temporal and concurrency properties of computations, communication and their synchronization (e.g. cyclic executive with time-triggered activation, priority-based pre-emption), which render a model statically analyzable. On the other hand, the synchronous approach~\cite{Halbwachs:2010} is suitable for the formal design and verification of reactive systems (e.g. flight control systems) that react to stimuli from the environment within strict time bounds. In synchronous reactive languages, the program reacts in a sequence of logical clock ticks and computations within a tick are instantaneous. In this setting, programs are amenable to formal verification~\cite{Hal99} and code generation for embedded platforms~\cite{1695922}.

The process network models of computation are mainly used in streaming signal processing and provide a means to cope with the complexity of parallel programming. They dictate a decomposition of behavior into pieces by defining the relationships between these pieces, while enabling the analysis of non-functional properties. Programs are written in the form of directed graphs with nodes for their functions and arcs for the data flows between functions. Such programs can exploit concurrency when they are deployed to parallel hardware, while their functions can be statically scheduled.

In~\cite{FASE-18}, we introduced the formal semantics of a new process network model that combines streaming and reactive control processing with task parallelism. The so-called FPPNs (Fixed Priority Process Networks) enable the design of applications that react to environment stimuli through the definition of communicating tasks that are programmed independently from the execution platform. Task activations depend on a combination of data availability (similar to streaming applications) and complex (non-periodic) arrival patterns. A noteworthy characteristic of FPPN-based designs is their functional determinism, i.e. the fact that for a given test stimuli we expect a deterministic output response. This feature renders applications amenable to testing, as opposed to applications programmed using low-level real-time OS services, whose outputs may depend on their timing behavior. The schedulability properties of FPPNs and the first static scheduling solutions were presented in~\cite{litDatePaper}, whereas their applicability in mixed-criticality systems~\cite{litISOLA} and in systems with shared resource interference~\cite{SEFM17} have been also studied.

The FPPN semantics is defined in~\cite{FASE-18} by compilation into RT-BIP~\cite{litRTBIP}, an executable formal language for modeling networks of connected timed automata. In present paper, we utilize the results from~\cite{FASE-18},~\cite{litDatePaper} and~\cite{litISOLA} to propose a design flow based on the principles of rigorous system design~\cite{EDA-034}. Such a flow differs from other model-based approaches in that a sequence of semantics-preserving steps allows deriving an executable implementation of an FPPN-based application design. This is achieved by embedding the functional code into the FPPN design model through the high-level architecture description interface of the TASTE toolset~\cite{Perrotin2012}, whose front-end tools have been amended to capture FPPN-compliant models. The schedulability is established by static analysis of the high-level FPPN, it is preserved throughout the subsequent model transformation steps and it is eventually guaranteed by construction by the derived implementation for the real-time BIP runtime environment that supports parallel execution of BIP components using POSIX threads. This means that the developers can reason in terms of high-level schedulabity concepts (e.g. tasks, priorities, deadlines, offsets etc.) and they do not need to explicitly program low-level real-time OS services (e.g. for task management, inter-task communication, memory allocation, scheduling etc.), thus retaining the predictability advantage of a statically analyzable design.

We experimented on the rigorous design of a Guidance Navigation \& Control (GNC) on-board spacecraft application that was ported onto ESA's Next Generation Microprocessor (NGMP), more specifically the quad-core LEON4FT processor~\cite{litLEON4FT}. The effectiveness of our approach is shown through measurements on execution traces, which also provide valuable insight for analyzing the scheduling bottlenecks.

In Section~\ref{sec:FPPN-TASTE-tool}, we introduce the basics of the FPPN model of computation and we then show the GNC FPPN model captured using the front-end tools of TASTE. Section~\ref{sec:rigorous} discusses in detail the design flow steps. Section~\ref{sec:schedulability} elaborates the steps of schedulability analysis and code generation, as well as the details of their application on the GNC case study. Section~\ref{sec:case-study} presents the measurements from the execution of the scheduled GNC FPPN C++ programs on top of the BIP runtime environment that was ported onto the LEON4FT processor. The paper concludes with a critical overview of the achievements, a discussion on the gained insight and the future research prospects.   

\section{Related model-based approaches for real-time embedded systems}

The entry point of any model-based design approach is an adequate language, which allows avoiding the premature commitment to specific execution/interaction semantics or implementation choices. In particular, the design of time-critical applications is a challenging procedure, which is usually based on evolutionary prototype building~\cite{Hugues:2008}. As the design cycle evolves, modeling formalisms are expected to support refinement, the setting of system attributes and analysis. The non-functional properties of the system are analyzed e.g. by model checking, simulation or other methods, and the results are taken into account in system design. In~\cite{BRAU20181}, the authors discuss and address two challenges related to the transformations of the model used for design into models used for analysis: the validity of the transformations and how to take the analysis result into account in system design.

Other related works~\cite{Radojevic:2011} emphasize the theory of model transformations and the incorporation of suitable models of computation, however, by often ignoring schedulability aspects. Model-based System-Engineering environments like Ptolemy II and PeaCE support a plethora of models of computation as a means for the refinement-based design of (multi-core) real-time embedded systems~\cite{Ha:2008,6864027,1173203}. To the best of our knowledge, this work presents the first model-based approach grounded on the principles of timing-aware rigorous design~\cite{litRTBIP} and task schedulability on multiple cores.

\section{FPPN process networks and their design with the TASTE toolset}
\label{sec:FPPN-TASTE-tool}

A process network model defines a partial order, in which a relative sequence of concurrent executions is specified only for a sequence of events, handled based on causality and inherent dependencies. While the engineer focuses on the key aspects of the application's design, in the physical implementation of every event it is somehow attached to a real-time instant, thus implying a natural total order for the system's execution. The engineers do not need to know the model's formal semantics; they just have to follow composition rules, which establish the dependencies between processes in the form of data and/or control flow.

In~\cite{FASE-18}, the authors proposed the FPPNs, a process network model that combines streaming and reactive control processing. An instance of FPPN is composed of three main entities: \emph{Processes}, \emph{Data Channels} and \emph{Event Generators}. A \emph{Process} represents a software subroutine, i.e. functional code that operates with internal variables and 
input/output channels connected to it through ports. Every process is mapped to an event generator, which determines whether the process is periodic or sporadic. Periodic and sporadic processes are a generalization of classical periodic and sporadic tasks to communication via channels. Processes are assigned \textit{functional priorities}, which define a relation between processes to ensure deterministic communication. An invocation of a process is referred to as a \emph{job}. Like the real-time jobs, the subroutine should have a bounded execution time subject to WCET (worst-case execution time) analysis.

An FPPN is formally defined by two directed graphs. The first one is a (possibly cyclic) graph $(P,C)$, whose nodes $P$ are processes and edges $C$ are channels for pairs of communicating processes that define a dataflow direction, i.e. from the writer to the reader (there are also external channels interacting with the environment). The second graph $(P, FP)$ is the functional priority directed acyclic graph (DAG) with edges defining a functional priority relation between processes. This part of FPPN's definition ensures its functional determinism, i.e. that the outputs calculated by FPPN depend
only on the event invocation times and the input data sequences, but not on the
scheduling.

\begin{figure}[!ht]
 \centering
 \includegraphics[width=0.7\linewidth]{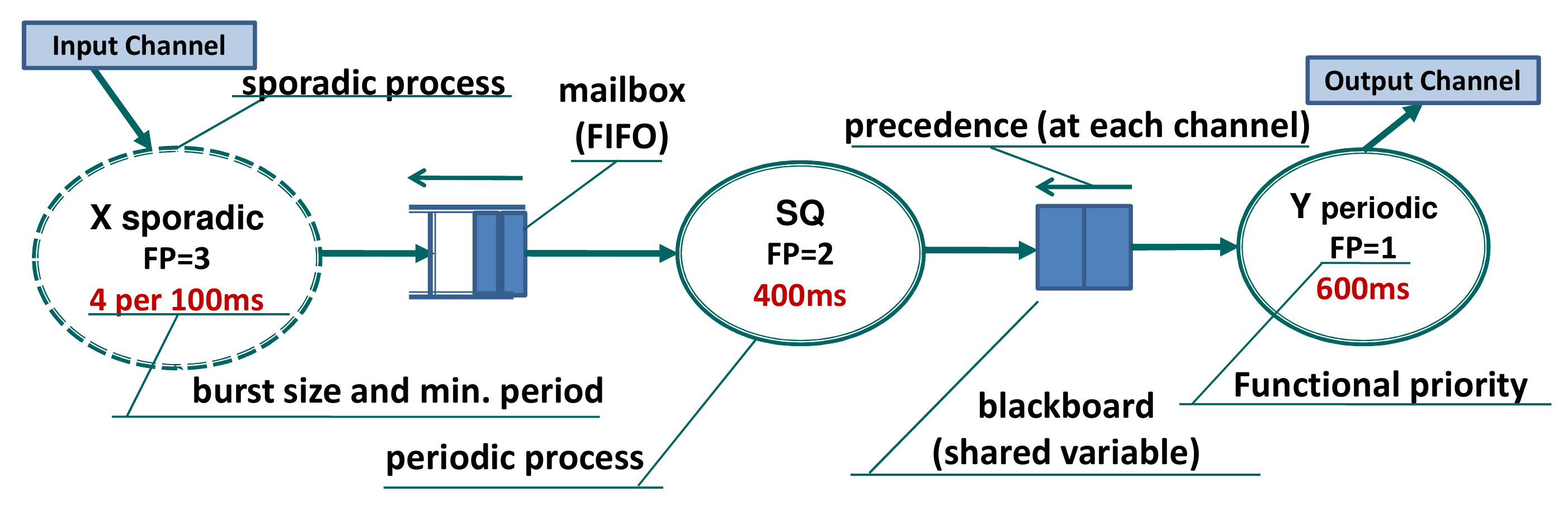}
 \caption{Example Fixed Priority Process Network}
 \label{figAppModel}
\end{figure}

The process network in Fig.~\ref{figAppModel}, represents a data processing application, where  the ``X'' sporadic process generates values, the ``Square'' process calculates the square of the received value and the ``Y'' periodic process serves as a sink for the squared value. A sporadic event (a command from the environment) activates ``X'' sporadic, which is annotated by its minimal inter-arrival time. The periodic processes are annotated by their periods. The two types of non-blocking inter-process channels are also illustrated. The FIFO (or mailbox) has a semantics of a queue. The blackboard remembers the last written value that can be read multiple times. The arc depicted above the channels indicates the functional priority relation $FP$ (higher to lower). 
Additionally, the environment input/output channels are shown. In this example, the dataflow in the channels goes in the opposite direction of the functional priority order.

To design FPPN models, we propose using the architecture description interface of the TASTE toolset~\cite{Perrotin2012}, whose front-end tools have been amended appropriately. The FPPN model is conceptualized and specified in the TASTE-IV (TASTE Interface-View) graphical editor, which generates an AADL (Architecture Analysis \& Design Language) syntax~\cite{litAADL} description of the graphical model representation.

In TASTE-IV, the TASTE functions are assigned attributes. The \texttt{FPPNClass} attribute defines the type of FPPN entities (e.g. blackboard, periodic process). The sporadic process is configured by coupling two TASTE functions: a function with \texttt{FPPNClass = sporadic} that provides a sporadic interface and a function with \texttt{FPPNClass=sporadic-protocol} that provides a periodic interface for polling for conditions to invoke the sporadic interface. To satisfy the precondition for the schedulability analysis~\cite{litDatePaper}, we assume that one and only one periodic process is connected to each sporadic process, i.e. there is always a single channel that connects a sporadic process to a periodic process. The FPPNClass attributes \texttt{mailbox} and \texttt{blackboard} are used for data-channels of the respective type; each channel declares two provided interfaces for `read' and `write', while the processes that access the channel have respective required interfaces. The \texttt{DataChannelSize} should be also defined, which represents the minimum size of the data type (in bytes) communicated via the channel. For a mailbox channel, the \texttt{DataChannelLength} is defined, which determines the length of the FIFO. 

The \texttt{Fpriority} attribute is an integer, which dictates the priority index of the process, and hence its priority order in the network. Each process is assigned a unique index. For any two processes, the one with the larger index has lower functional priority than the other. 

In the TASTE-DV (Data View) editor, the designer specifies in ASN.1 language format the types of exchanged data via FPPN channels. This specification is independent from the implementation languages and platforms, and allows deriving automatic marshallers that follow any kind of binary encoding rules. 

C code skeletons are generated from the TASTE I-V model together with an XML file that specifies the system topology. The functional code of the application under design employs FPPN templates that reflect in C language the primitives of FPPN processes and channels. The designer can use the TASTE Deployment-View for compiling, binding and running the functional code of the application on his workstation development environment. A first evaluation of the application's functionality is thus possible, without having scheduled its tasks according to their functional priority order and their real-time constraints.

\begin{example} \label{ex:GNC-FPPN}
A Guidance Navigation \& Control (GNC) on-board spacecraft application controls the movement of the vehicle by processing the data of the corresponding sensors and controllers. GNC involves three steps: the guidance equipment and software first compute the orbital location required to satisfy mission requirements, the navigation then tracks the vehicle's actual location, and the flight control directs the orbit to the required location. Depending on the specific phase, some components can be inactive, and/or the specific data to be exchanged can have a different format. For example, during the orbital phase, the guidance function will provide inertial reference attitude to the controller, whereas in the re-entry phase reference aerodynamic angles will be sent. The application comprises:

\begin{compactitem}
\item The \textbf{Guidance Navigation Task} that estimates the current translational state of the vehicle based on measurements by the sensors and on actuator commands, whenever applicable. This task keeps the vehicle on track during the flight to reach the desired location for parachute triggering. It calculates the actual location and provides the reference attitude and the calculated air data and aerodynamic parameters to the control task, which in turn assures its objective. If the reference attitude is pre-computed (e.g. coming from reference trajectory for the Orbital phase), it will also pass through this block to keep the function's definition general. This is a periodic process with period $T_p$ = 500ms, deadline $d_P$ = 500ms and worst-case execution time $C$=22ms.
\item The \textbf{Control FM task} that runs the control and flight management algorithms. This is a periodic process with period $T_p$ = 50ms, deadline $d_P$ = 50ms and worst-case execution time $C$=8ms. 
\item The \textbf{Control Output Task}, which sends the outputs of the GNC (geodetic altitude, longitude, mach and dynamic pressure) to the Dynamics Kinematics and Environment module. This is a periodic process with period $T_p$ = 50ms, deadline $d_P$ = 50ms and worst-case execution time $C$=4ms.
\item The \textbf{Data Input Dispatcher Task}, which reads, decodes and dispatches data to the right destination whenever new data from the spacecraft's sensors are available. In our GNC model the data input dispatcher processes MVM (Mission and Vehicle Management), IMU (Inertial Measurement Unit) and GPS (Global Positioning System) data, which have been pre-computed through measurements by their relevant subcomponents and are stored in C buffers. This is a periodic process with period $T_p$ = 50ms, deadline $d_P$ = 50ms and worst-case execution time $C$=6ms.
\end{compactitem} 
  
Fig.~\ref{fig:GNC-MSC} delineates the exchange of information between GNC tasks in a single hyperperiod of 500ms (least common multiple of all periods). The FPPN model in Fig.~\ref{fig:GNC-FPPN} was designed based on the Message Sequence Chart and the source code, for identifying how the different tasks interact by conditional signals. The functional priorities of the FPPN impose precedence from numerically smaller (higher-priority) to numerically larger. Fig.~\ref{fig:GNC-FPPN-APP} depicts the TASTE-IV FPPN model for the GNC application.

\begin{figure}[!ht]
\centering
\includegraphics[scale=0.6]{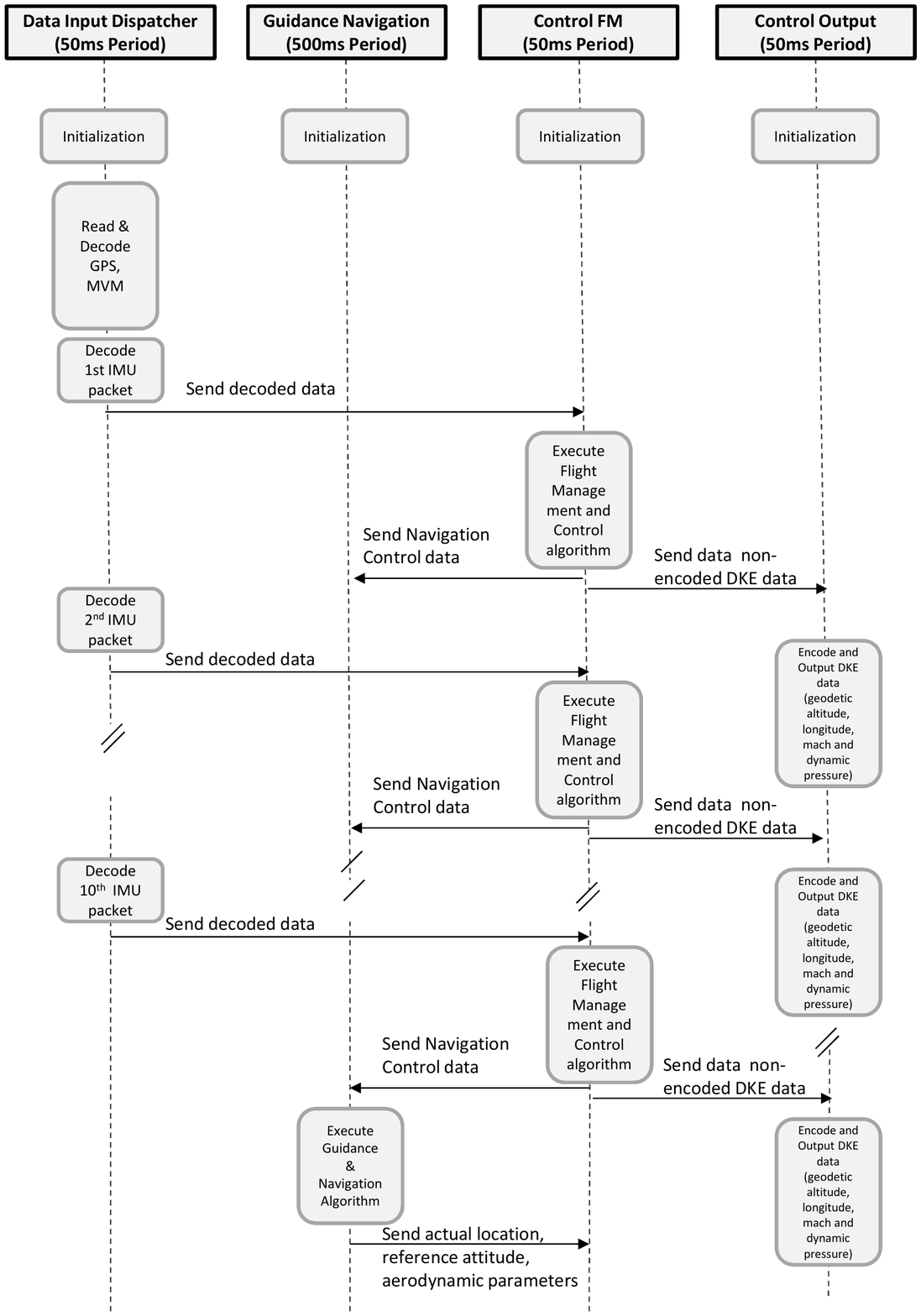} 
\caption{The Message Sequence Chart of the GNC application}
\label{fig:GNC-MSC}
\end{figure}

\begin{figure}[!ht]
\centering
\includegraphics[scale=0.5]{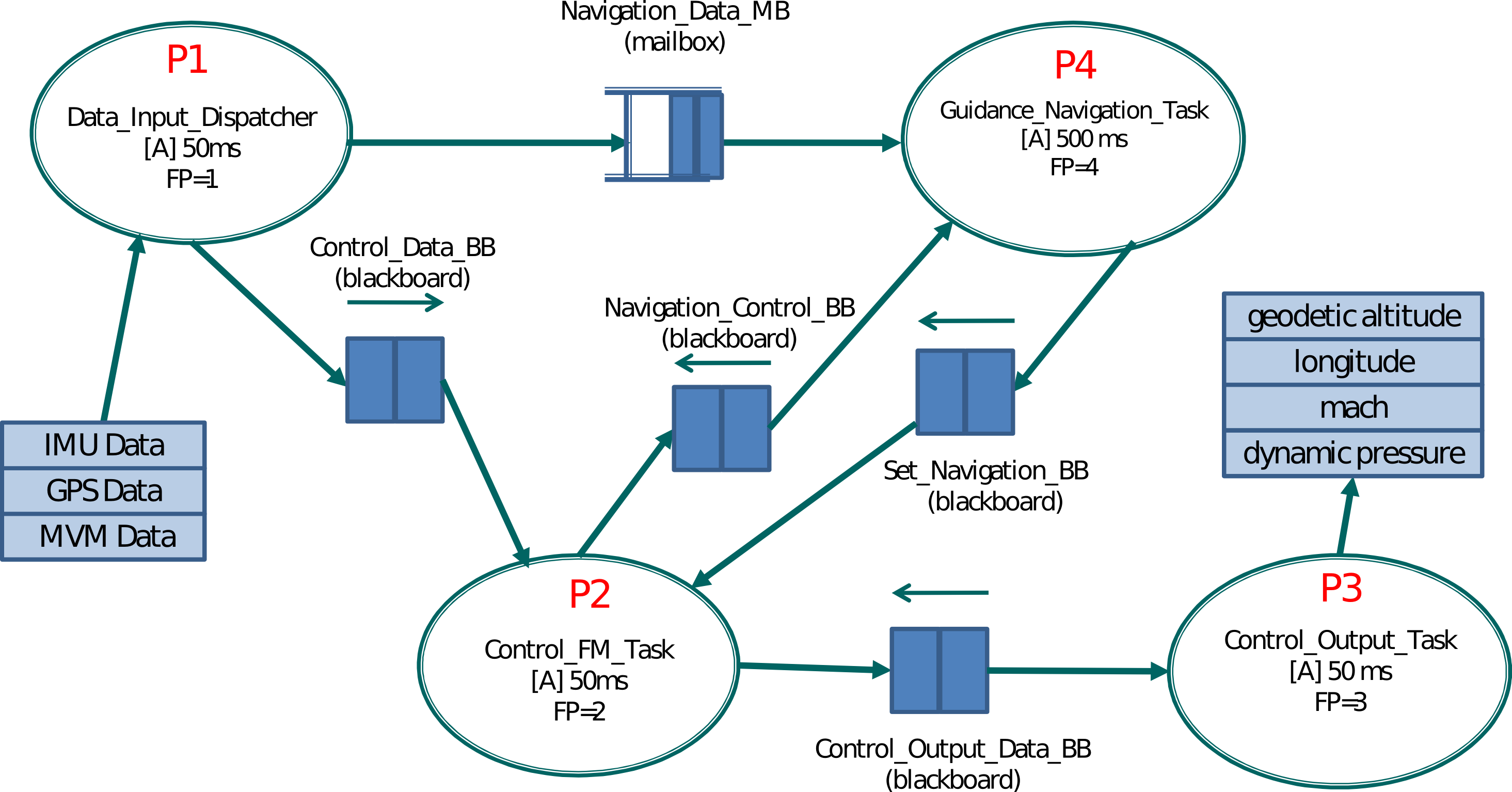} 
\caption{The GNC FPPN model}
\label{fig:GNC-FPPN}
\end{figure}

\begin{figure}[!ht]
\centering
\includegraphics[scale=0.17]{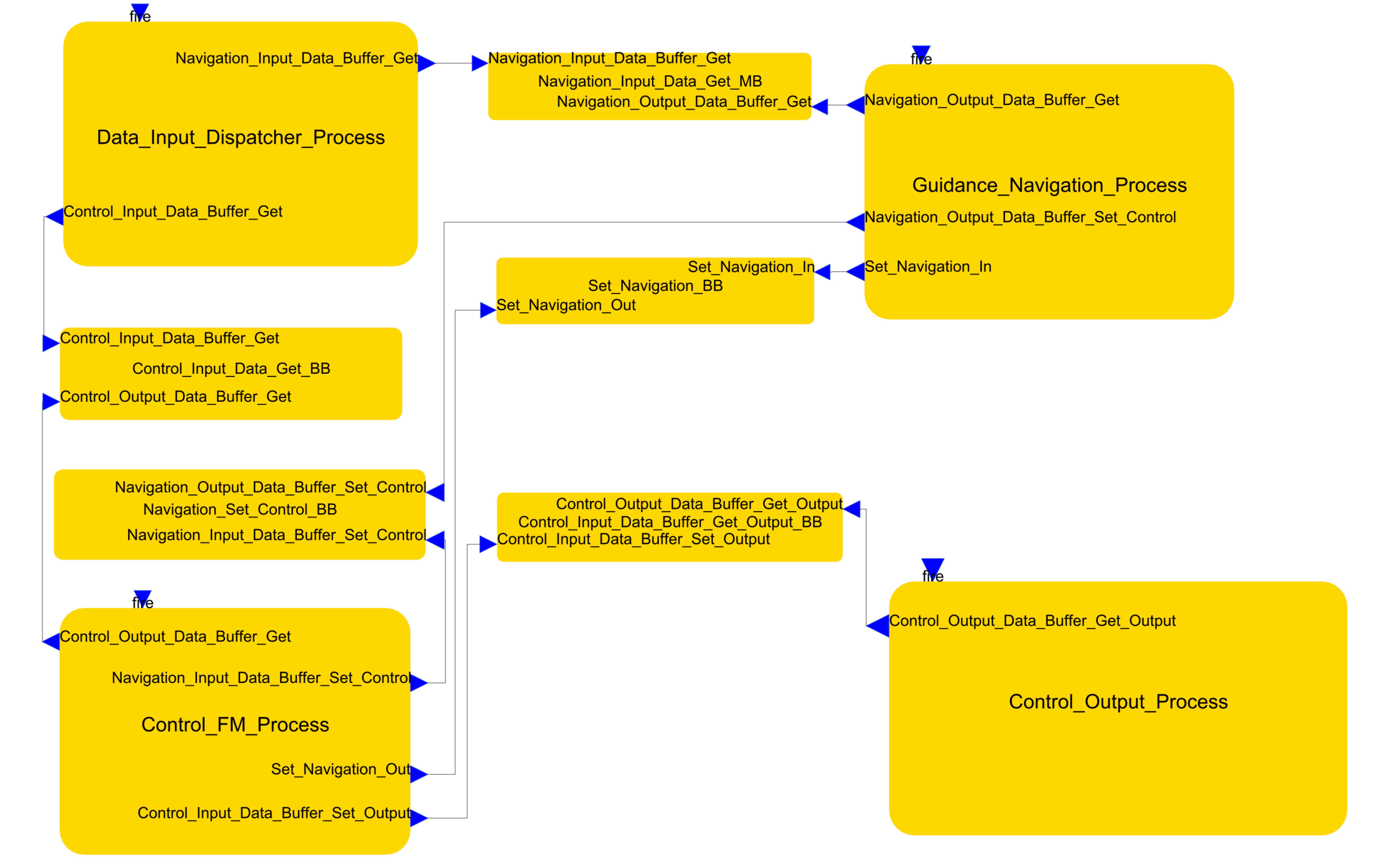} 
\caption{The TASTE-IV FPPN model for the GNC application}
\label{fig:GNC-FPPN-APP}
\end{figure}

\end{example}

\section{Rigorous design of embedded real-time systems based on FPPNs}
\label{sec:rigorous}

The proposed design approach aims to a formal and accountable process for deriving implementations of FPPNs whose schedulability is preserved throughout the process and it is eventually guaranteed by construction on top of the real-time BIP execution engine. A series of automated model transformation steps is applied starting with the TASTE2BIP\footnote{TASTE2BIP is online in: \url{http://www-verimag.imag.fr/Time-Critical-Applications-on-Multicore}} transformation~\cite{FASE-18} that compiles any FPPN process network into the BIP language. The source code is parsed, searching for primitives (reads and writes from/to the data channels) that are relevant for the process interactions. In addition to the BIP models, the TASTE2BIP model transformation also derives a task graph for static scheduling, as detailed in~\cite{litDatePaper}.

\begin{definition}[Task Graph]
A directed acyclic graph $TG(J,E)$ whose nodes $J=\{J_i\}$ are jobs defined by tuples $J_i = (p_i,k_i,A_i,D_i,C_i)$, where
  $p_i$ is the job's process,
  $k_i$ is the job's invocation count,
  $A_i \in Q_{\ge 0}$ is the arrival time,
  $D_i \in Q_+$ is the absolute deadline and
  $C_i \in Q_+$ is the WCET.
The $k$-th job of process $p$ is denoted by $p[k]$. The edges $E$ represent constraints on the job execution order.
\end{definition}

\begin{example}
Fig.~\ref{fig:task-graph} depicts the task graph derived from the GNC FPPN model of Example~\ref{ex:GNC-FPPN}.
\begin{figure}[!ht]
\centering
\includegraphics[width=.8\textwidth]{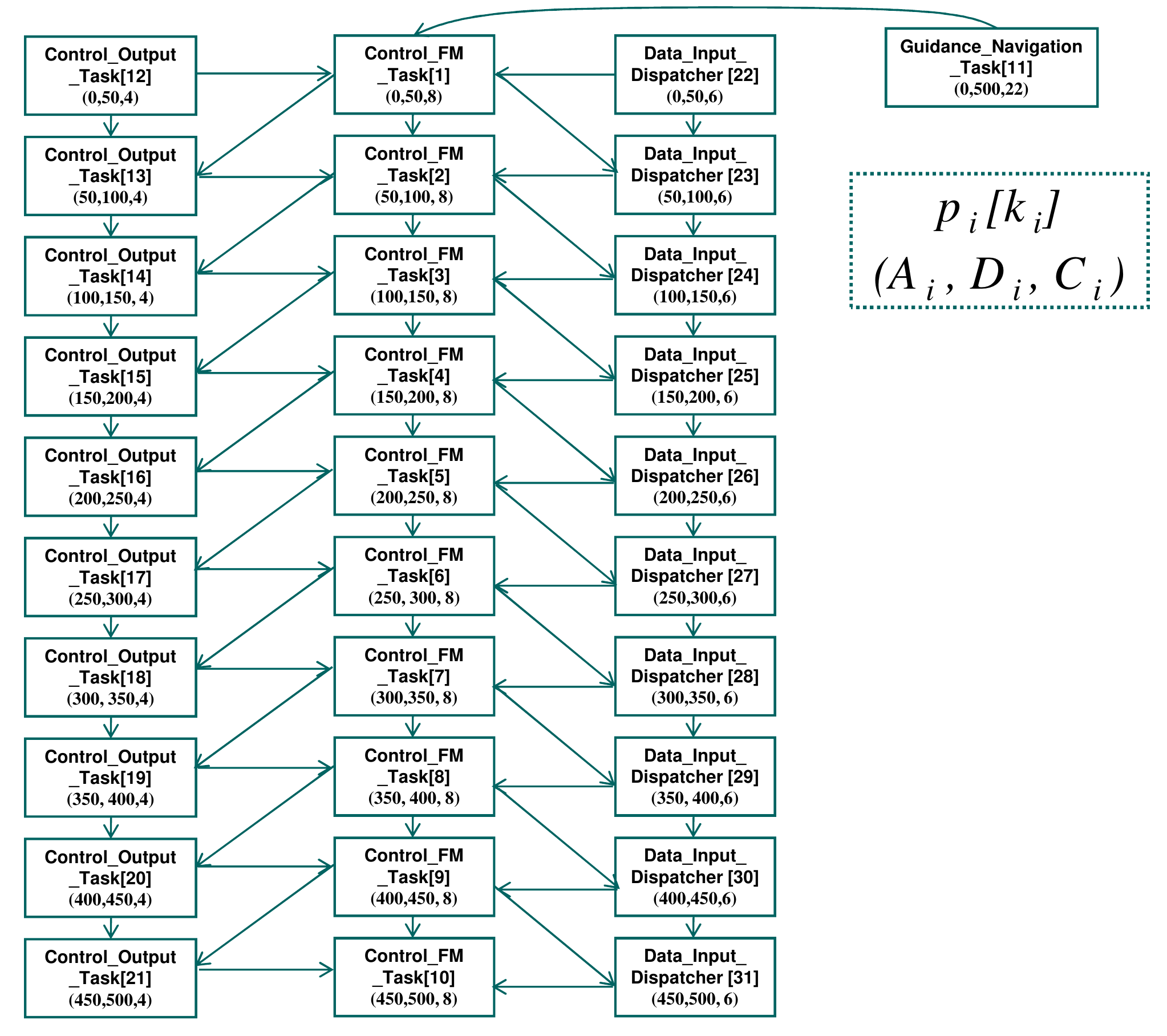} 
\caption{Task graph for the GNC FPPN model}
\label{fig:task-graph}
\end{figure}
\end{example}

\begin{figure}[!ht]
\centering
\includegraphics[width=.93\textwidth]{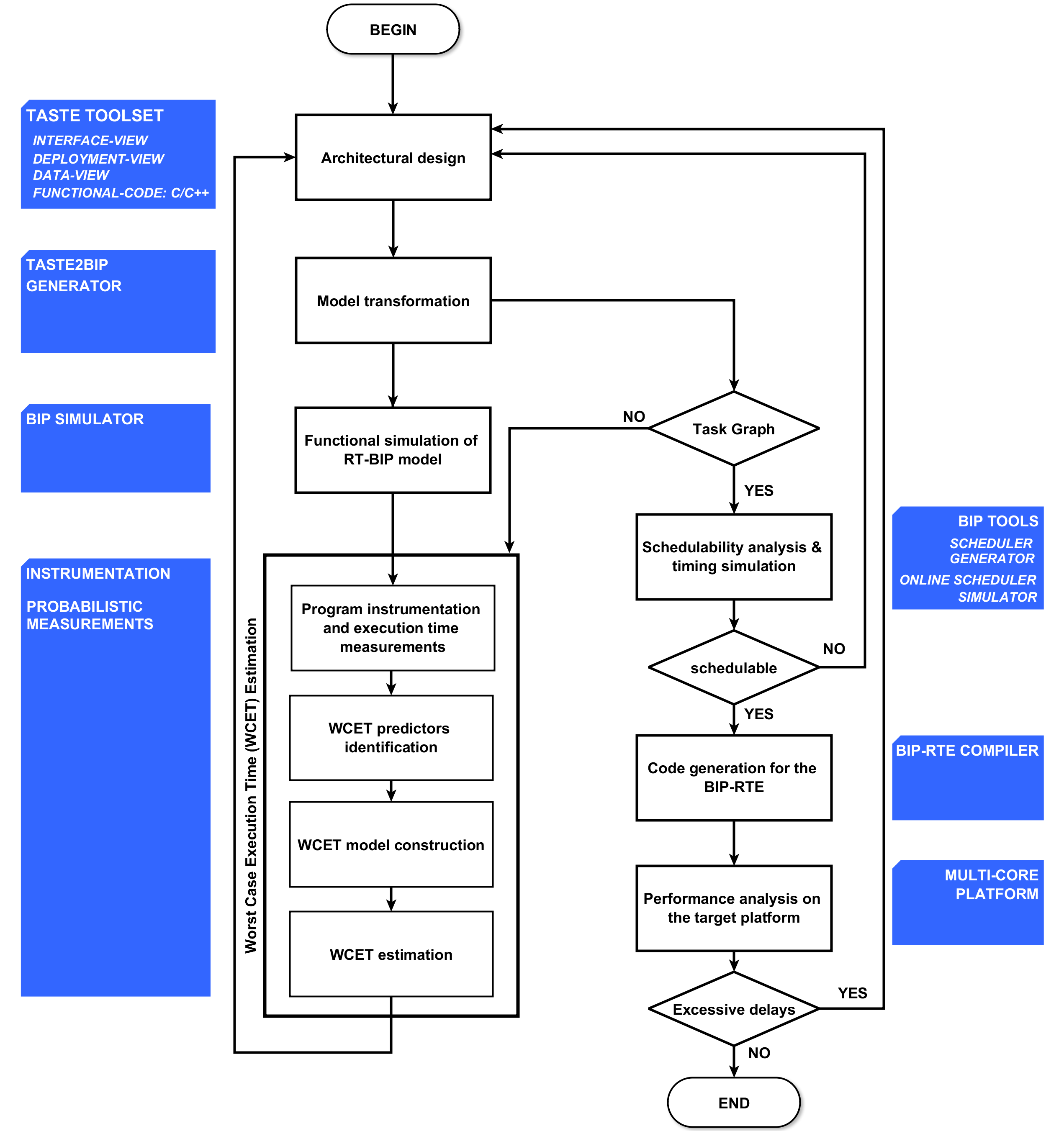} 
\caption{Rigorous design flow for embedded systems based on the real-time BIP execution engine}
\label{fig:design-flow}
\end{figure}

Figure~\ref{fig:design-flow} shows the design flow steps and the tool-support associated with each step. More precisely, the design flow develops as follows.

\begin{compactitem}
\item[\textit{Input}:] (i) application requirements (SW items, real-time constraints) \\
(ii) platform requirements (HW platform \& resources)
\item[\textit{Output}:] implementation on the target platform

\item[Step 1] \textit{Architectural design:} The functional code (software behaviour) is implemented and the requirements are mapped to an architectural model (i.e., TASTE I-V). This encompasses the static architecture (static decomposition into software components) and the dynamic architecture (active objects such as threads, tasks and processes, along with their resource and I/O dependencies). 

\item[Step 2] \textit{Model transformation:} FPPN model transformation into RT-BIP according to the FPPN execution semantics in~\cite{FASE-18}. If WCETs are known, the task graph is also generated.   

[\emph{if (Task Graph exists) goto Step 5}]

\item[Step 3] \textit{Functional simulation of RT-BIP model:} The processor time requirements of the application are judged on the basis of the BIP model's execution on the target platform (Step 4); before this, the same model should be functionally tested on a workstation.
\item[Step 4] \textit{Worst Case Execution Time (WCET) Estimation: The probabilistic measurement-based timing analysis in~\cite{PopNAZBK17,Nouri2018} is used that can arguably guarantee safe probabilistic bounds.} 
\begin{compactitem}
\item[4.1] \textit{Program instrumentation: A trace point is inserted at every branching of control flow of the task code. The program is run on a workstation to collect execution traces, which are used to identify all potential predictors.}
\item[4.2] \textit{WCET predictors identification: A sufficient subset of the set of potential predictors is identified for an adequate execution time regression model.}
\item[4.3] \textit{WCET model construction: The Maximal Regression Model for conservative overestimation of the execution time is applied~\cite{PopNAZBK17}.}
\item[4.4] \textit{WCET estimation: A Maximal Execution Time bound is computed.} 

[\emph{goto Step 1}]
\end{compactitem}

\item[Step 5] \textit{Schedulability analysis \& timing simulation:} The task graph is given as input to a static scheduler, along with the attributes of each job (the process to which the job belongs, its invocation count, arrival time, absolute deadline, WCET).

[\emph{if (! schedulable) iterate Steps 1 to 4}]

\item[Step 6] \textit{Code generation for the BIP RTE:} The joint application/scheduler model is compiled by the RT BIP compiler and linked with the BIP-RTE.

\item[Step 7] \textit{Performance analysis on the target platform:} Validation by performance analysis is essential towards identifying possible excessive delays, due to resource starvation cases.
 
[\emph{if (excessive delays found) goto Step 1}]

\end{compactitem}

Apart from the TASTE2BIP transformation, functional simulation in step 3 takes place using the RT-BIP tools. The statistical tools for the WCET estimation in step 4 are described in~\cite{PopNAZBK17}. In step 5, the schedule obtained from the scheduler is translated into input for the online-scheduler model in BIP, which constraints job executions for resource management (task to processor core mapping and other constraints). In steps 4.3 and 7, the executable runs on the target platform on top of the real-time operating system (RTEMS-SMP). For the analysis in step 7, appropriate tools are used (e.g. gprof) that trace/monitor the software performance on the target platform.

\section{Schedulability analysis and code generation for the BIP RTE}
\label{sec:schedulability}

\begin{figure}[!ht]
\centering
\includegraphics[width=.7\textwidth]{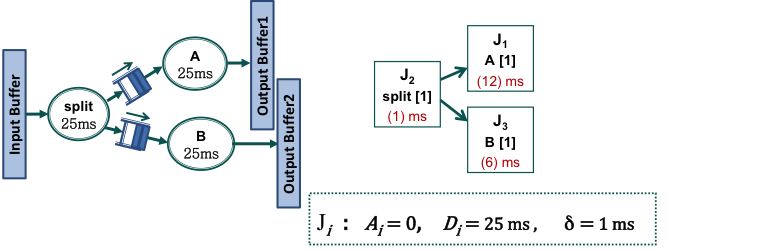} 
\caption{The FPPN model of a system with three tasks}
\label{fig:three-tasks}
\end{figure}

To illustrate the main principles of FPPN scheduling we consider the synthetic application in Fig.~\ref{fig:three-tasks} with three tasks. The ``split'' task appends two small data items (a few bytes) to the two output channels and sleeps for 1 ms to imitate some task execution time. Tasks ``A'' and ``B'' read the data and Task “A” sleeps for 12 ms whereas Task ``B'' sleeps for 6 ms. All tasks have the same periodic scheduling window, with period and deadline being 25 ms. In a real application, this corresponds to the time during which the two input data buffers should be read, the computations performed and the output buffers written. In the derived task graph, every task is represented by a job. The jobs are numbered as $J_i = J_1, J_2, J_3$ and annotated by their WCETs. The arrival times $A_i$ and deadlines $D_i$ for all jobs are the same.

In step 5 of the design flow, the static scheduler accepts a parameter $\delta$ for the worst-case cost of a single transition in the BIP automata components. Parameter $\delta$ is platform-dependent and characterizes the coarse-grain interference between the task components, when they access the centralized BIP RTE engine to execute discrete automata transitions. In the simplest possible scheduling, every job is executed as soon as it arrives and its predecessors have finished (ASAP policy). In this case, the functional priority ordering (\texttt{Fpriority} attributes in the TASTE-IV FPPN) that is implemented in the generated RT-BIP model enforces the predecessor - successor relation between the jobs thus ensuring deterministic updates to the channel states.

\begin{figure}[!ht]
\centering
\includegraphics[width=.73\textwidth]{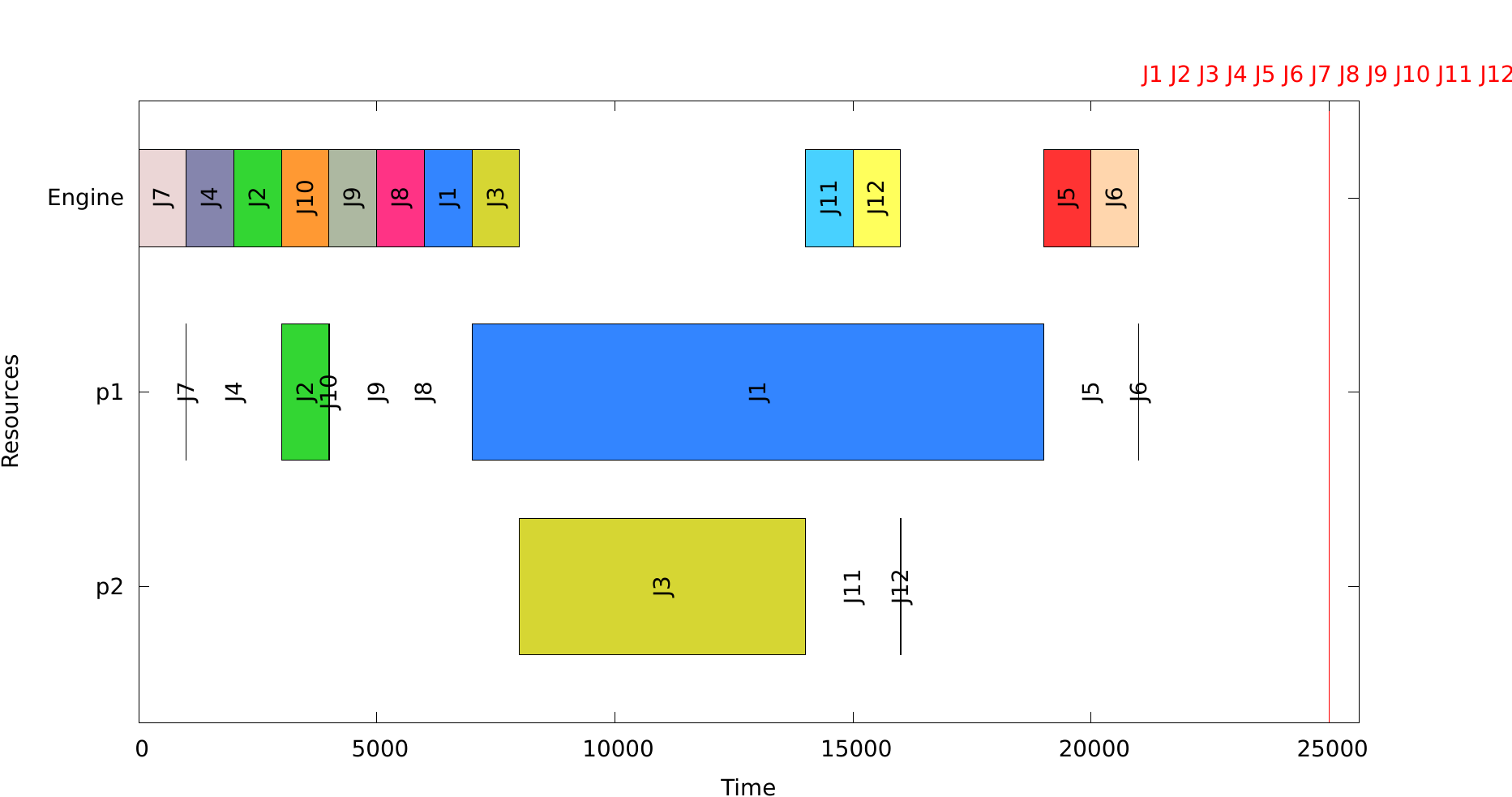} 
\vspace{-9pt}
\caption{FPPN schedule computed by the static scheduler for the three-tasks example}
\label{fig:three-tasks-schedule}
\end{figure}

The scheduler tool produces the time-triggered table depicted in Fig.~\ref{fig:three-tasks-schedule} with the times when discrete BIP transitions occur, which are imposed by the online scheduler. The blocks executed on Core 0 correspond to discrete BIP transitions of duration $\delta=1 ms$. 

The tool applies list scheduling based on an heuristically computed priority relation (consistent with the functional priority relation), a total order in which earlier jobs have higher priority. This involves a simple simulation of the fixed-priority policy~\cite{litDatePaper}. For the task graph of Fig.~\ref{fig:three-tasks} two compute cores are needed. This happens due to the 12 ms interference overhead (four BIP transitions required per task's execution, take 4$\cdot \delta =$ 4 ms); the task graph cannot be scheduled on a single processor, because an amount 12+(1+12+6) ms = 31 ms of processor time per period of 25 ms has to be allocated. In the schedule of Fig.~\ref{fig:three-tasks-schedule}, Task “split” and Task “A” are mapped to Core 1 and Task “B” to Core 2.

An expressive timed automata language like BIP provides potential means to implement custom strategies for controlling interference on shared HW/SW resources. As we have seen in the above example, interference has to be taken into account together with the WCET in the schedulability model. It can be controlled through adoption of an appropriate interference model, as shown by the authors of~\cite{litISOLA}. In that work, additional schedulability perspectives are also considered, most notably mixed-criticality scheduling.

\section{The GNC application with BIP RTE running on a quad-core platform}
\label{sec:case-study}

Through the rigorous design flow we had convenient means to experiment with various scheduling scenarios when running the GNC application with BIP RTE on a LEON4FT embedded platform. Our aim was to explore the speedup in throughput potential for each scenario. We present here the results for a ``pipelined'' scenario which represented an attempt to exploit multi-core parallelism as much as possible; the potential for such an exploratory approach is inherent in the FPPN model and remains transparent to the application designer until the final steps of our rigorous design flow.       

Specifically, we focused on the functional priority relation of Fig.~\ref{fig:GNC-FPPN}, where the Control Output (P3) and Guidance Navigation (P4) tasks process the data received in the previous period. Therefore, they both have enough data to start execution immediately at the beginning of the period. Thus, the priority relation, which should normally for this application follow in Fig.~\ref{fig:GNC-FPPN} the direction from left to right, for the channels connected to these two tasks, follows the opposite direction. Note that the mailbox channel even has no functional priority arrow associated with it, which will be justified below. Thus, P3 and P4 have no predecessors, and the same holds for P1, and the three of them can run in parallel. The absence of arrow for the mailbox between P1 and P4, which makes it possible for them to run in parallel, is explained as follows. Normally the data channels should be protected by priority arrow to avoid data races, but instead we increased the capacity of the mailbox to make a double buffer out of it. Thus, the reader process, P4, can read one data item, whereas the writer process, P1, writes the other one, and they can do it concurrently without risking data races (but with risking some interference, as we see later on). 

\begin{figure}[!ht]
\centering
\includegraphics[width=.75\textwidth]{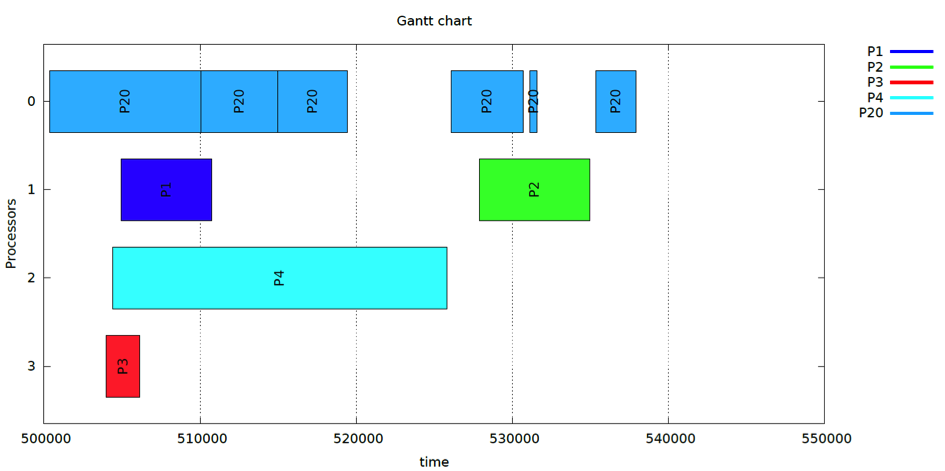} 
\caption{Zoom in a time window with parallel job executions of the GNC on the LEON4FT processor}
\vspace{-9pt}
\label{fig:gnc-leon4b}
\end{figure}

Fig.~\ref{fig:gnc-leon4b} shows the Gantt chart, which zooms in the time-window, where the jobs are executed in parallel on the 3 cores plus the fourth core (core 0) where the BIP RTE engine executes the BIP transitions.

We see that all activities in the current period end by time slightly less than 40 ms after the period start. This gives us a (reverse proportional) measure of a maximal throughput that the system could achieve, by reducing the period from 50 ms to 40 ms. Despite a higher parallelism compared to a scenario where three cores are utilized~\cite{FASE-18}, no improvement is achieved in throughput due to the interference between P1, P4, which manifested in longer execution time of P4. A possible solution to this end is modifying the mailbox BIP implementation to reduce the interference.

\section{Conclusion}

We introduced a rigorous approach for multi-core embedded system design based on the FPPN process network model and the BIP RTE. FPPNs allow reasoning in terms of high-level schedulability concepts, whereas predictability on the execution platform is guaranteed by construction, since the BIP RTE enables the consistent mapping of user-programmed scheduling policies to operating system mechanisms (e.g. threads, dynamic priorities).

\nocite{*}
\bibliographystyle{eptcs}
\bibliography{generic}
\end{document}